\documentclass[aps,prb,twocolumn,superscriptaddress,showpacs]{revtex4-1}

\usepackage{graphicx,amsmath}

\usepackage{amssymb}

\begin{document}

\title{Anisotropic two-orbital Hubbard model: single-site versus cluster dynamical
mean-field theory}

\author{Hunpyo Lee}
\affiliation{Institut f\"ur Theoretische Physik, Goethe-Universit\"at Frankfurt, Max-von-Laue-Stra{\ss}e 1, 60438 Frankfurt am Main, Germany}
\author{Yu-Zhong Zhang}
\affiliation{Institut f\"ur Theoretische Physik, Goethe-Universit\"at Frankfurt, Max-von-Laue-Stra{\ss}e 1, 60438 Frankfurt am Main, Germany}
\affiliation{Department of Physics, Tongji University, Shanghai, 200092 P. R. China}
\author{Harald O. Jeschke}
\affiliation{Institut f\"ur Theoretische Physik, Goethe-Universit\"at Frankfurt, Max-von-Laue-Stra{\ss}e 1, 60438 Frankfurt am Main, Germany}
\author{Roser Valent\'\i}
\affiliation{Institut f\"ur Theoretische Physik, Goethe-Universit\"at Frankfurt, Max-von-Laue-Stra{\ss}e 1, 60438 Frankfurt am Main, Germany}

\date{\today}

\begin{abstract}
  The anisotropic two-orbital Hubbard model with different bandwidths
  and degrees of frustration in each orbital is investigated in the
  framework of both single-site dynamical mean-field theory (DMFT) as
  well as its cluster extension (DCA) for clusters up to four sites
  combined with a continuous-time quantum Monte Carlo algorithm.  This
  model shows a rich phase diagram which includes the appearance of
  orbital selective phase transitions, non-Fermi liquid behavior as
  well as antiferromagnetic metallic states. We discuss the advantages
  and drawbacks of employing the single-site DMFT with respect to DCA
  and the consequences for the physical picture obtained out of these
  calculations. Finally, we argue that such a minimal model may be of
  relevance to understand the nature of the antiferromagnetic metallic
  state in the iron-pnictide superconductors as well as the origin of
  the small staggered magnetization observed in these systems.
\end{abstract}


\maketitle

\section{Introduction}

The multiorbital Hubbard Hamiltonian has been
proposed~\cite{Imada1998} to be a suitable model for the description
of a large variety of correlated electron transition metal compounds
including nickelates, vanadates, and more recently iron pnictide and
chalcogenide superconductors~\cite{Kamihara08}.  While new interesting
phases like orbital-selective metal-insulator phase transitions,
coexistence of paramagnetic with antiferromagnetic phases or
coexistence of Fermi liquid with non-Fermi liquid behavior have been
predicted for this model, there is still a strong debate regarding the
true nature of the phase diagram.  The reason for that lies in the
complexity of the multiorbital model which includes orbital degrees of
freedom, crystal field splitting effects as well as the Hund's rule
exchange coupling and also in the nature of the approaches used
in order to investigate the model.  In this work, we shall present a
comparative investigation of the properties of the anisotropic
two-orbital Hubbard model within the single-site dynamical mean field
approximation (DMFT)~\cite{Metzner1989,Georges1996,Kotliar2004} as
well as the dynamical cluster approximation (DCA)~\cite{Maier2005}
where spatial fluctuations are included.

We will focus on three major phenomena related to the multiorbital
Hubbard model. (i) Orbital selective phase transitions (OSPT) which
have been discussed, for instance, in
Ca$_{2-x}$Sr$_{x}$RuO$_4$~\cite{Anisimov2002,Fang2004} and recently in
iron-based superconductors~\cite{Lee2010(3),Oh2010} in the framework
of DMFT and density functional theory (DFT) calculations. Within
single-site DMFT it has been shown that the presence of OSPT is
dependent on various effects like the existence of inequivalent
bandwidths at integer filling~\cite{Knecht2005,Liebsch2004,Arita2005,Koga2004},
the presence of crystal field splitting at non-integer
filling~\cite{Wener2007}, the absence of interorbital
hybridization~\cite{Medici2005} as well as the appearance of different
number of degenerate orbitals~\cite{Medici2009}.  Nevertheless, it is
unclear how these features are modified if spatial fluctuations are
included in the calculations. For instance, the single-site DMFT
cannot capture the Slater physics caused by nesting effects as is the
case in the spin-Peierls system TiOCl
~\cite{Saha-Dasgupta2005,Saha-Dasgupta2007} where dimerization is
important.
%and in photoemission spectra of iron~\cite{Schafer2005} where a strong momentum dependence of effective mass renormalization is observed.
(ii) Appearance of non Fermi-liquid (FL) behavior.  Non-FL behavior is
present in the single-site DMFT calculations for the
multiorbital~\cite{Liebsch2005,Biermann2005} Hubbard model and in the
four-site cellular-DMFT (CDMFT) studies for the one-orbital Hubbard
model~\cite{Zhang2007,Park2008}. Based on the mentioned calculations,
the mechanism of non-FL behavior is attributed to the ferromagnetic
correlation between orbitals (Hund's coupling) in the case of
multiorbital systems, while intersite antiferromagnetic correlations
caused by the Coulomb repulsion are suggested as the driving mechanism
in the one-orbital case. In an attempt to consider the effect of
spatial fluctuations in the multiorbital Hubbard model, we
investigated recently the mechanism of non-FL
behavior~\cite{Lee2010(1),Lee2010(2)} within DCA.  This study is
developed further in the present work.  (iii) Finally, we investigate
the competing effect of frustrated orbitals in a multiorbital Hubbard
model.  Such a study has been motivated by the intensively discussed
controversy about the origin of the observed reduced magnetic moment
in Fe-based superconductors~\cite{Lee2010(3)} and the possible
relation to the multiorbital nature of the system.  Bandstructure
calculations together with downfolding for various families of
iron-based SC compounds show that weakly frustrated $d_{xy}$,
$d_{z^2}$ and highly frustrated $d_{yz/zx}$, $d_{x^2-y^2}$ orbitals
coexist~\cite{Miyake09,private_comm}. We will show in what follows
that competition of frustrated with non-frustrated orbitals induces an
antiferromagnetic metallic phase with a reduced ordered magnetic
moment~\cite{Lee2010(3),Lee2010(2)}, as observed in iron-based
superconductors.

In section~\ref{model} we present the model and numerical methods
considered which include the single-site DMFT, DCA and the
continuous-time QMC algorithm as impurity solver. In
section~\ref{results}, we discuss the appearance of orbital selective
phase transitions and non-Fermi liquid behavior for two models: ({\bf I})
the half-filled two-orbital Hubbard model with unequal bandwidths on
the square lattice in the framework of a four-site DCA paramagnetic
(PM) solution and ({\bf II}) the half-filled two-orbital Hubbard model on the
Bethe lattice with one frustrated orbital and one unfrustrated orbital
using the single-site DMFT with AF ordering.  In this model we find an
antiferromagnetic metallic phase as well as a reduced staggered
magnetization as a consequence of competing frustration effects.
Finally, in section~\ref{summary} we summarize our findings.

\section{Model and Methods\label{model}}

We consider the anisotropic two-orbital Hubbard model with nearest-
and next-nearest- hopping matrix elements:
\begin{eqnarray}
  H &=&-\sum_{\langle ij\rangle m\sigma}  t_m c^{\dagger}_{im\sigma}
  c_{jm\sigma} -\sum_{\langle ij'\rangle m\sigma}  t_m^{\prime} c^{\dagger}_{im\sigma}
  c_{j'm\sigma} \nonumber \\
    &&+\,U\sum_{im}n_{im\uparrow} n_{im\downarrow} +\sum_{i\sigma\sigma'}(U'-\delta_{\sigma \sigma'} J_z)n_{i1\sigma}
  n_{i2\sigma'},\label{eq:H}
\end{eqnarray}
where $c_{im\sigma}(c^{\dag}_{im\sigma})$ is the annihilation
(creation) operator of an electron with spin $\sigma$ at site $i$ and
orbital $m$. $t_m$ ($t_m^{\prime}$) is the hopping matrix element
between site $i$ and nearest-neighbor site $j$ (next-nearest-neighbor
site $j'$). $U$ and $U'$ are intraorbital and interorbital Coulomb
repulsion integrals, respectively, and $J_z n_{i1 \sigma}n_{i2
  \sigma}$ is the Ising-type Hund's rule coupling term. In our
calculations we set $U'=\frac{U}{2}$ and $J_z=\frac{U}{4}$ in order to
fulfill the condition $U=U'+2J_z$ and evaluate the model with the
single-site DMFT and four-site DCA methods combined with a
continuous-time QMC algorithm.

Even though DCA can be considered as superior to the single-site DMFT
method, both approaches are faced with certain advantages and
drawbacks. While the single-site DMFT ignores spatial fluctuations, a
notorious fermionic sign problem present in the QMC calculations
disappears even in the frustrated systems with next nearest-neighbor
hopping $t'$. On the other hand, while short-range spatial
fluctuations are considered in the four-site DCA approach, the QMC
calculations encounter a serious fermionic sign problem for the
frustrated model~\cite{Gull2008(1)}. For these reasons, we first consider the anisotropic two-orbital model
with different bandwidths $t_m$ and no frustration $t'_m$ = 0 on the
square lattice at half-filling within the four-site DCA method (model
({\bf I}) above).  Although a mild fermionic sign problem with
average sign around $\langle s\rangle=0.98$ is observed, we believe
that the results are numerically reliable.

In the DCA method the reduced Brillouin zone (BZ) is determined by
dividing the total BZ with respect to the number of sites and the
self-energies for the different cluster momenta are calculated --
analogous to the DMFT equation -- on the reduced BZ by considering the
same assumption as in DMFT where the self-energy is constant in the
(reduced) Brillouin zone.  The DCA self-consistent equations for each
cluster momentum with ${\bf K}=(0,0),(0,\pi),(\pi,0)$ and $(\pi,\pi)$
are given as
\begin{eqnarray}\label{DCAE}
\overline{G}_{\sigma}({\bf K},i\omega_n)=\frac{1}{N}\sum_{\tilde{\bf K}}
\frac{1}{i\omega_n+\mu-\epsilon_{{\bf K+\tilde{K}}}-\Sigma_{\sigma}({\bf K},i\omega_n)},
\end{eqnarray}
where $N$ is the number of sites, $\mu$ is the chemical potential,
$\epsilon_{{\bf K+\tilde{K}}}$ is the dispersion relation, $\omega_n$
is the Matsubara frequency, and the summation over ${\bf \tilde{K}}$
is calculated in each reduced Brillouin zone.

Next, we will study the anisotropic frustrated two-orbital Hubbard
model on the Bethe lattice at half-filling with $t_m$ and $t'_m$
hopping matrix elements (model ({\bf II}) above) by employing the
single-site DMFT with an antiferromagnetic bath.  The N\'eel state, in
which the magnetization of A and B sublattices are in opposite
directions, is included in the model and the DMFT self-consistency
equations are given as
\begin{eqnarray}
 G_{0,A,\sigma}^{-1}=i\omega_n + \mu - t_m^2 G_{B,\sigma} - t_m^{\prime 2}G_{A,\sigma}, \\
 G_{0,B,\sigma}^{-1}=i\omega_n + \mu - t_m^2 G_{A,\sigma} - t_m^{\prime 2}G_{B,\sigma},
\end{eqnarray}
The sign problem is absent in the single-site impurity calculations,
and the converged impurity Green's functions in both cases are
determined after several iterations.

Two possible continuous-time (diagrammatic) QMC algorithms for solving
the DMFT and DCA equations have been proposed.  One is based on an
expansion in the impurity-bath hybridization
$\Delta$~\cite{Werner2006(1),Werner2006(2)} term -- also called
strong-coupling approach -- and the other one is based on an expansion
in the interaction term $U$~\cite{Rubtsov2005,Lee2008,Assaad2007},
denoted as weak-coupling approach.  Unlike a determinant QMC
method~\cite{Hirsch1986} which includes the artificial Trotter error
caused by discretization of the imaginary time, this error is not
present in the continuous-time QMC algorithms except as numerical
noise. In addition, since the matrix sizes of both diagrammatic QMC
methods are smaller than that of the determinant QMC method, it is
possible to access the low temperature regime.  For our study we
employ the weak-coupling QMC method. The reason for this choice lies
in the computational effort. While the strong-coupling QMC method for
the single-site impurity problem of multiorbital systems has 
advantages like the possibility of considering the full Hund's rule
coupling term without any fermionic sign problem and also the
calculation at very low temperatures in strong-coupling regions, the
numerical effort of such an approach increases rapidly with the number
of cluster sites $N$ as $2^{2N}$.  Therefore, in the multiorbital
cluster systems the strong-coupling QMC method is computationally
extremely expensive.  On the other hand, the weak-coupling QMC method
contains less numerical burden because the matrix size increases only
linearly with $N$.

\section{Results\label{results}}
\subsection{Orbital selective phase transitions and non-Fermi liquid
  behavior in the four-site DCA calculations}

In this section, we present results which demonstrate the appearance
of OSPT and non-FL behavior in the half-filled two-orbital Hubbard
model with different hopping $t_m$ in each orbital $m$ on the square
lattice (model ({\bf I})) by employing the four-site DCA calculations
with paramagnetic bath.  The hopping $t_m$ of the narrow and wide
bands are $t_1=0.5$ and $t_2=1.0$, respectively, and the
temperature considered is $T/t_2=0.1$.
\begin{figure}[h!tb]
\centerline{
\includegraphics[width=0.8\linewidth]{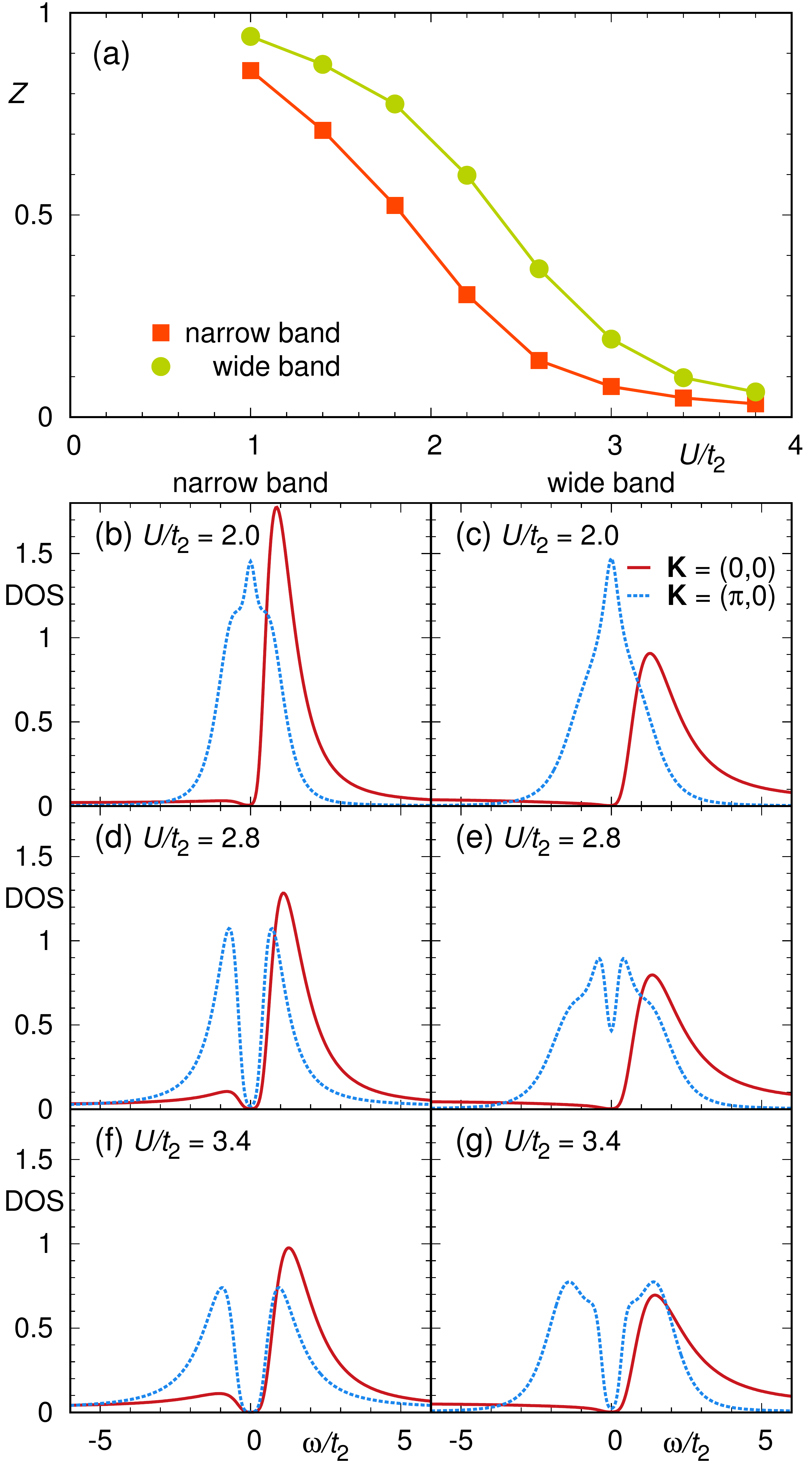}}
\caption{\label{FigLayer1} (a) The quasiparticle weight $z_{\bf K}$ at
  ${\bf K} = (\pi,0)$ as a function of $U/t_2$ for the narrow and wide
  bands. The spectral functions at $T/t_2=0.1$ for (b)-(c)
  $U/t_2=2.0$, (d)-(e) $U/t_2=2.8$, and (f)-(g) $U/t_2=3.4$ in the
  four-site dynamical cluster approximation (DCA) calculation.}
\end{figure}
In order to investigate the existence of the OSPT, we first explore
the quasiparticle weight $z_{\bf K}$ as a function of $U/t_2$, which
is given as $z_{\bf K} \approx \big[1-\frac{\text{Im} \Sigma({\bf K},
    i\omega_0)}{\omega_0}\big]^{-1}$, where $\omega_0$ is the lowest
Matsubara frequency and ${\bf K}$ is the momentum sector.  The
effective mass ratio, which is defined as $z_{{\bf K}}^{-1}$, gives
information on correlation effects. As is well known, correlation is
the driving force for a transfer of spectral weight from the coherent
peak at the Fermi level ($\omega / t_2 = 0)$ to incoherent peaks.
This means that the quasiparticle weight $z_{\bf K}$ at the Fermi
surface ${\bf K}=(\pi,0)$ should be suppressed in a Mott insulating
state due to enhancement of the effective mass.  The results are shown
in Fig.~\ref{FigLayer1}~(a). The quasiparticle weight $z_{\bf K}$ at
${\bf K} = (\pi,0)$ in the narrow band is suppressed around
$U/t_2=2.8$, while it remains large in the wide band.  We can denote
this feature as an orbital-selective phase where metallic and
insulating states coexist between $U/t_2=2.8$ and 3.2. In order to
observe the OSPT more clearly at momentum sectors ${\bf K} = (0,0)$
and $(\pi,0)$, we plot in Figs.~\ref{FigLayer1}~(b)-(g) the spectral
functions for $U/t_2=2.0$, $2.8$ and $3.4$, where the maximum entropy
method was used for analytical continuation.  The spectral functions
at ${\bf K} = (0,0)$ show a band-insulator, while at ${\bf K} =
(\pi,0)$ they indicate metallic states in the weak-coupling region
$U/t_2=2.0$ in both bands like in results obtained from
non-interacting four-site DCA in the single-band Hubbard
model~\cite{Haule2007}.  Upon increasing the interaction we find that
the spectral functions indicating a band insulator at ${\bf K} =
(0,0)$ are almost unchanged.  On the other hand, we observe that for
$U/t_2=2.8$ the gap is completely opened at ${\bf K} = (\pi,0)$ in the
narrow band, while a finite density of states around the Fermi level
remains in the wide band, clearly showing the coexistence of an
insulator and a metal in this two-orbital system as displayed in
Figs.~\ref{FigLayer1}~(d)-(e).  In the strong-coupling region with
$U/t_2=3.4$ in Figs.~\ref{FigLayer1}~(f)-(g), both phases at ${\bf K}
= (\pi,0)$ are insulating and the gap in the narrow band is larger
than that in the wide band. This means that between $U/t_2=2.8$ and
$3.4$ a Mott metal-insulator transition at ${\bf K} = (\pi,0)$ occurs
in the wide band, while the narrow band is already in the insulating
state; at the same time, a band insulator is sustained at ${\bf K} =
(0,0)$, regardless of the strength of the interaction $U/t_2$.

\begin{figure}[h!tb]
\centerline{
\includegraphics[width=0.9\linewidth]{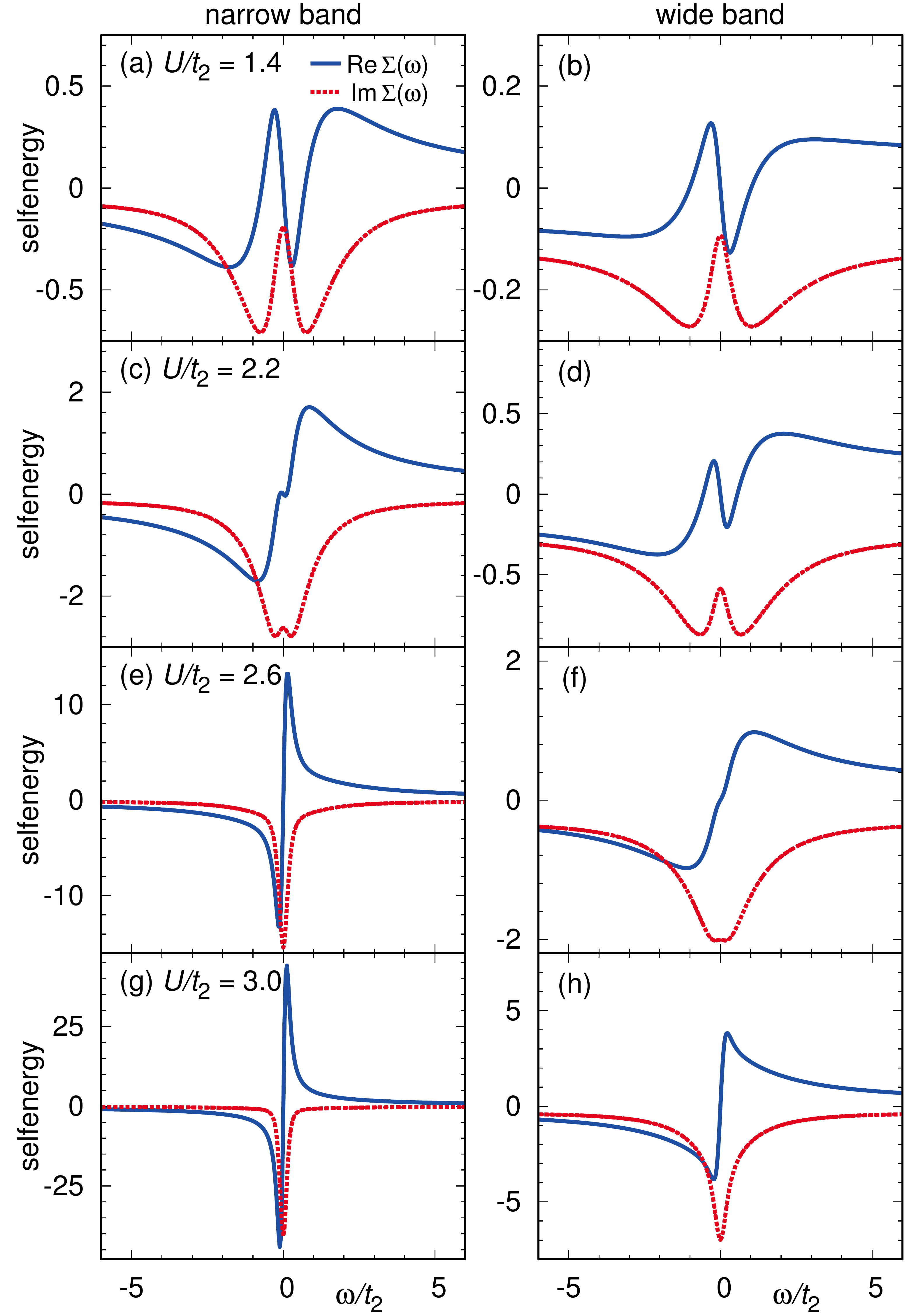}}
\caption{\label{FigLayer2} Self-energies at the Fermi surface ${\bf K}
  =(\pi,0)$ as a function of real frequency $\omega / t_2$ at $T/t_2=0.1$ for the narrow and wide bands
  in the four-site DCA calculations: (a) and (b)
  $U/t_2=1.4$, (c) and (d) $U/t_2=2.2$, (e) and (f) $U/t_2=2.6$, and
  (g) and (h) $U/t_2=3.0$.}
\end{figure}

Next, we plot the self-energies as a function of real frequency in
Figs.~\ref{FigLayer2}~(a)-(h) to show the FL, non-FL and Mott
insulating behavior clearly.  The analytical continuation of the
self-energy suggested by Wang et al.~\cite{Wang2009} has been
used. Here we briefly introduce the method: at high frequency
$\omega_n$ the self-energy at half-filling should behave as
$\Sigma(i\omega_n) \sim \frac{U^2}{4 \omega_n}$.  In order to perform
the Fourier transformation of $\Sigma(i\omega_n)$, we need to take
into account the correction of the high frequency tail as
\begin{eqnarray}
 \Sigma(\tau) \sim \frac{1}{\beta} \Big(\sum_{n=-N_h}^{N_h} e^{-i\omega_n \tau}\Sigma(i\omega_n)  + \sum_{n=-\infty}^{\infty} e^{-i\omega_n \tau} \frac{4}{U^2\omega_n} \nonumber \\ 
-\sum_{n=-N_h}^{N_h} e^{-i\omega_n \tau} \frac{4}{U^2\omega_n}\Big),
\end{eqnarray}
where $\frac{1}{\beta} \sum_{n=-\infty}^{\infty} e^{-i\omega_n \tau^{+}}
\frac{1}{i\omega_n}=-0.5$ and $N_h$ is the highest number of Matsubara
frequencies.  Here, $\Sigma(\tau)$ is inserted into the maximum
entropy equations to obtain $\text{Im} \Sigma(\omega)$, where $\omega$
is the real frequency. $\text{Re} \Sigma(\omega)$ is then calculated
by the Kramers-Kronig relation.

In the weak-coupling region, $U/t_2=1.4$, Figs.~\ref{FigLayer2}
(a)-(b) show that the imaginary part of the self-energy approaches
zero as $\omega \rightarrow 0$ for both orbitals indicating FL
states. As the interaction is increased, a non-FL state in the narrow
band coexists with a FL state in the wide band as shown in
Figs.~\ref{FigLayer2}~(c)-(d). By further increasing $U/t_2$, the
narrow band shows a Mott insulating behavior where $\text{Im}
\Sigma(\omega)$ diverges at $\omega =0$ (Fig.~\ref{FigLayer2}~(e)),
while the wide band is still metallic with non-FL behavior (see
Fig.~\ref{FigLayer2}~(f)).  In the strong-coupling region $U/t_2=3.0$
in Figs.~\ref{FigLayer2}~(g)-(h), divergences of the imaginary part of
the self-energy in both orbitals indicate a complete Mott insulating
state.

\subsection{OSPT and small magnetic moment in the single-site DMFT calculation}

\begin{figure}[h!tb]
\centerline{
\includegraphics[width=0.8\linewidth]{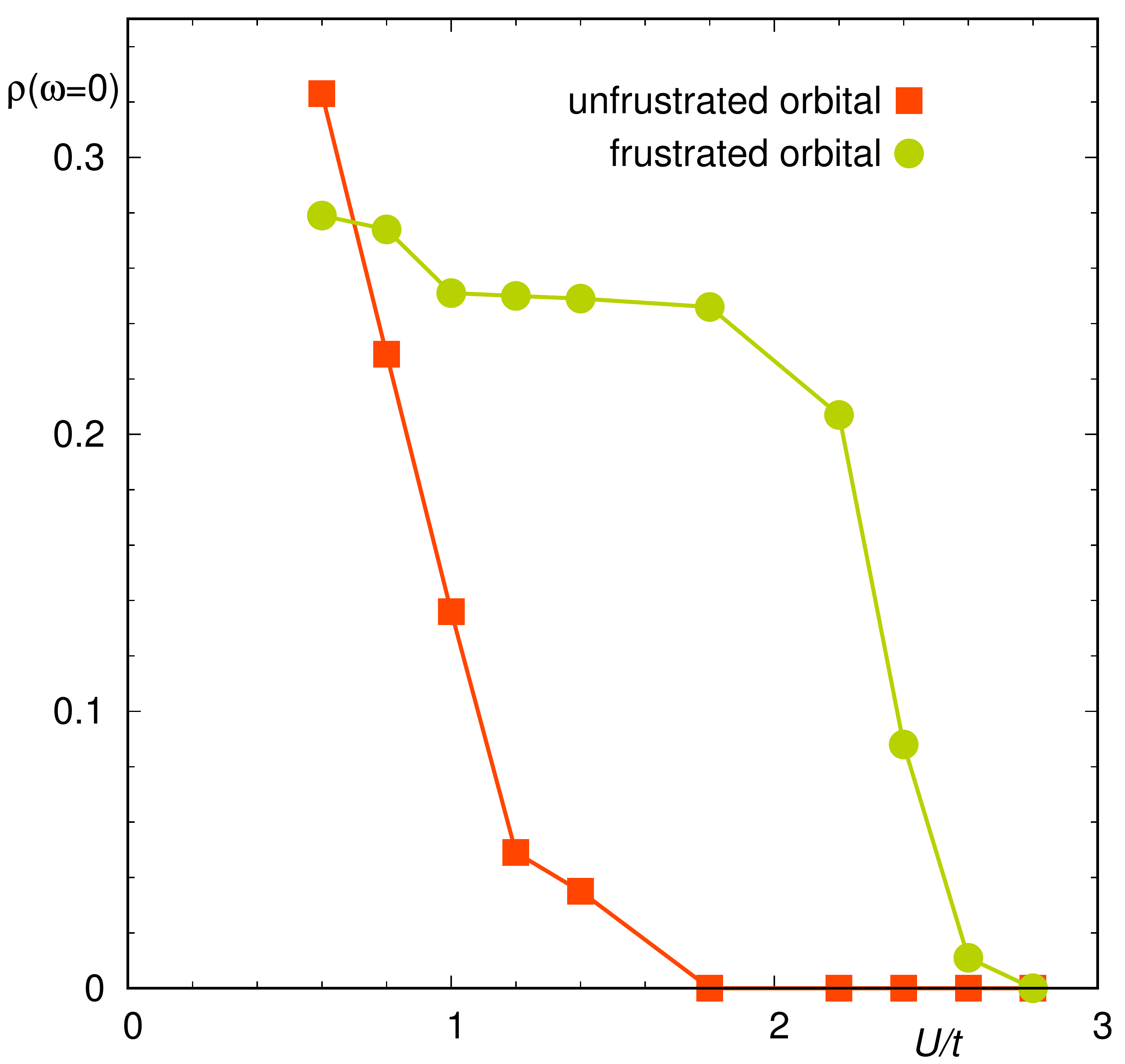}}
\caption{\label{FigLayer3} The density of states $\rho(\omega = 0)$ at
  the Fermi level as a function of $U/t$ for the unfrustrated and
  frustrated orbitals in the single-site dynamical mean field theory
  (DMFT) calculations.}
\end{figure}

We proceed now to the analysis of the anisotropic two-orbital Hubbard
model with frustrated and unfrustrated orbitals at half-filling using
the two-sublattice single-site DMFT with antiferromagnetic bath.  In
our calculations, the temperature is set to $T/t=1/16$. The next
nearest neighbour hopping is $t'/t = 0$ for the unfrustrated orbital
and $t'/t=0.65$ for the frustrated orbital with $t=1.0$.  We first
evaluate the OSPT behavior of this model with the single-site DMFT, as
we did within the four-site DCA approach in the former section.  We
observe that a band insulator is present since the source of the gap
opening is not only the Coulomb interaction but also the magnetization
of the AF solution.  Therefore, the change of the effective mass
-- which is related to the quasiparticle weight $z$ -- is very small in
contrast to the case of a Mott insulator.  Therefore, we employ the
DOS at Fermi level $\rho(\omega=0)$ to identify the OSPT as is shown
in Fig.~\ref{FigLayer3}. The results were obtained through
analytical continuation applying the maximum entropy method
and we performed around $10^7$
QMC samplings and  checked the results through a Pade approximation
in order to ensure the reliability of the results.
 The gap opening is clearly observed around
$U_{c_1}/t=1.9$ in the unfrustrated orbital, while the metallic state
is still present in the frustrated orbital.  As the interaction is
increased, a metal-insulator transition in the frustrated orbital
appears around $U_{c_2}/t=2.9$. Between
$U_{c_1}/t < U/t < U_{c_2}/t$ we find a coexistence region of
metallic and insulating states as a clear evidence of OSPT behavior.
Comparing our results of Fig.~\ref{FigLayer3} with those obtained with
the single-site DMFT with a paramagnetic solution in the two-orbital
Hubbard model with different bandwidths~\cite{Knecht2005}, we find
that in both cases the spectral densities $\rho(\omega=0)$ decay
suddenly to zero in the narrow (unfrustrated) band and the values in
the wide (frustrated) bands are similar around $U_{c_1}$.  The
difference between these two results is that $\rho(\omega=0)$ for the
frustrated orbital of our system also drops to zero at $U_{c_2}$,
while it decreases smoothly in the single-site DMFT calculation with
paramagnetic solution~\cite{Knecht2005}.  We believe that this is due
to the different mechanisms of the gap opening in each system.

The present model is also relevant for understanding the magnetic
behavior of the iron-pnictide superconductors.  Since the discovery of
high-$T_c$ superconductivity in iron pnictides in
2008~\cite{Kamihara08}, a lot of studies have been performed to find
the mechanism of pairing in these
materials~\cite{Mazin2008(1),Kuroki2008,Hirschfeld2009}.  Unlike the
high-$T_c$ cuprate superconductors, whose parent compounds are
antiferromagnetic insulators, an ordered antiferromagnetic metallic
state with small magnetic moment is detected in iron pnictides.
Density functional theory calculations on iron pnictides overestimate
the values of the ordered magnetic moment compared with experimental
results~\cite{Mazin2008(2),Opahle2009,Zhang2009,Zhang2010}.  Only by
artificially considering a negative $U$ or by employing a specific
double counting correction within
LDA+$U$~\cite{Ferber2010,Cricchio2010} can these values be reduced .
Even though some DFT+DMFT calculations including quantum fluctuations
have been performed, the mechanism of the small magnetic moment is
still
controversial~\cite{Yildirim2008,Si2008,Han2009,Schmidt2010,Haule2008,Craco2008,Haule2009,Skornyakov2009,Ishida2010,Bascones2010,Yang2010,Misawa2010,Yin2010}.
Here, within model ({\bf II}) we discuss a possible mechanism for the
small magnetic moment in iron pnictides and the presence of a metallic
antiferromagnetic state.

\begin{figure}[h!tb]
\centerline{
\includegraphics[width=0.85\linewidth]{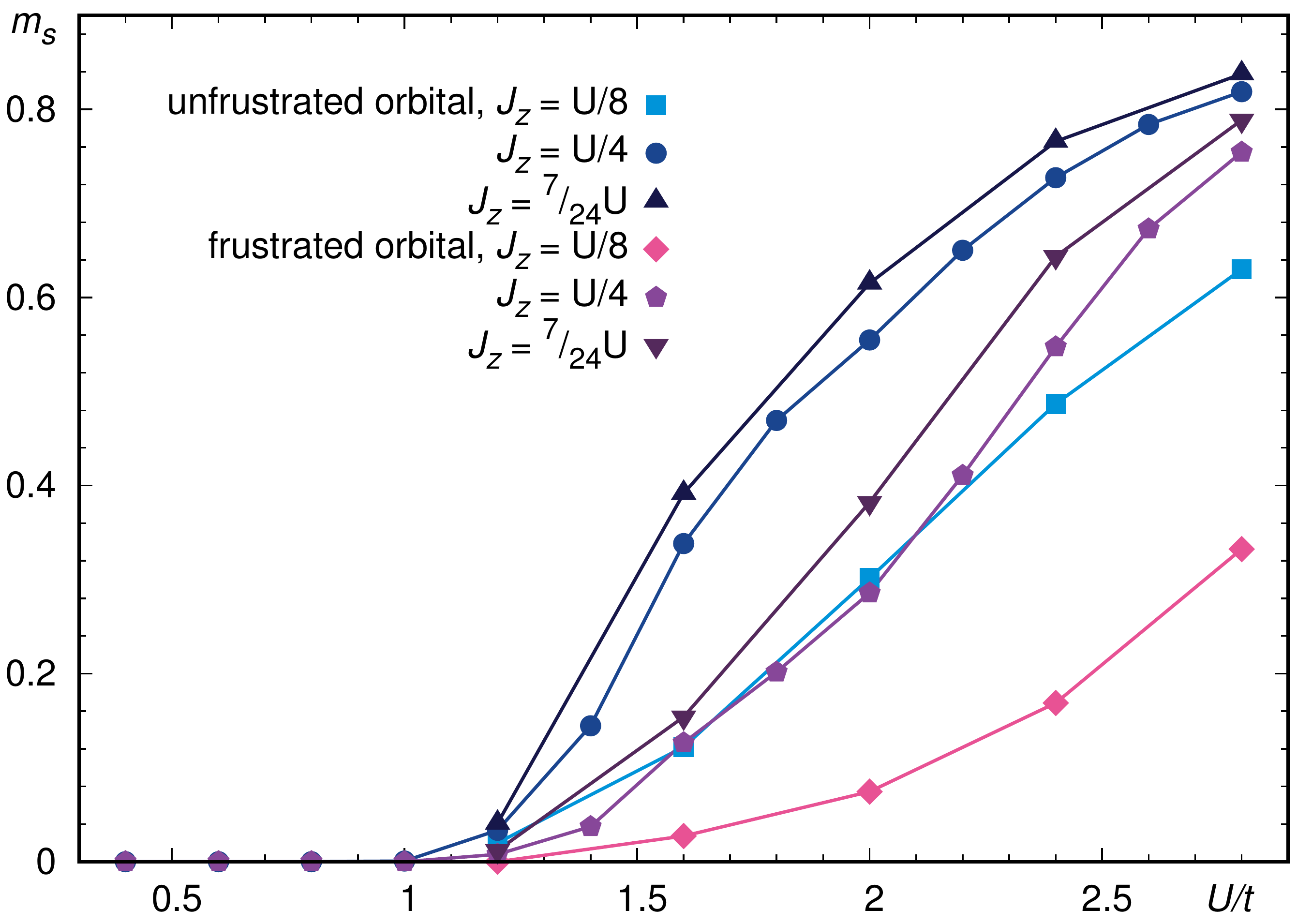}}
\caption{\label{FigLayer4} Magnetization as a function of $U/t$ with
  different values of $Jz$ in the single-site DMFT calculations.}
\end{figure}

In our former studies, we proposed the coupling between frustrated and
unfrustrated orbitals in a multiorbital Hubbard model as the possible
mechanism for the small magnetic moment observed in iron
pnictides~\cite{Lee2010(3)}.  Based on the minimal two-orbital Hubbard
model presented above, we found~\cite{Lee2010(3)} a continuous metal
to insulator phase transition as well as a continuous magnetic phase
transition for an Ising-Hund's rule coupling of $J_z=U/4$.  The
reduced ordered magnetic moment is shown in Fig.~\ref{FigLayer4} as a
function of $U/t$. We observe at a critical $U/t$ a smooth increase of
the magnetic moment, which is in contrast to the first order phase
transition behavior observed in the one-orbital frustrated Hubbard
model~\cite{Zitzler2004}. In order to show that the results are
independent of the choice of $J_z$ we also show in
Fig.~\ref{FigLayer4} the magnetization for $J_z=7U/24$, $U/8$ and
$U/4$ with the constraint of $U=U'+2J_z$.  As the interaction is
increased, the magnetization in all cases increases smoothly and in
the intermediate interaction regime small magnetizations are observed.

\section{Summary\label{summary}}

In the present work we have explored the properties of two
models. ({\bf I}) The two-orbital Hubbard model with unequal
bandwidths on the square lattice by employing the four-site DCA and
({\bf II}) the two-orbital Hubbard model with a frustrated and an
unfrustrated orbital on the Bethe lattice by using the single-site
DMFT method.  The DCA and DMFT equations were solved using a
continuous-time QMC algorithm. The above choice of systems and methods
explicitly avoids the fermionic sign problem in the QMC
calculations. We find that both models show a rich phase diagram as a
function of interaction strength which includes the appearance of
orbital selective phase transitions, non-Fermi liquid behavior as well
as antiferromagnetic metallic states.

Using the four-site DCA method with a PM solution for the two-orbital
Hubbard model with different bandwidths (model ({\bf I})) we find
that while the system behaves as a band insulator in the K=(0,0) sector
of the Brillouin zone, it
undergoes  an orbital selective phase transition in  the K=($\pi$,0) sector.
The system in the K=($\pi$,0) sector shows FL behavior for both orbitals in the weak-coupling region.
As the interaction is increased, a non-FL state for the narrow band
coexists with a FL behavior of the wide band.  In the
intermediate-coupling regime the non-FL behavior is changed into a
Mott insulating state where the electrons are completely localized in
the narrow band, while the metallic state with non-FL behavior is
still observed in the wide band.  In the strong-coupling region the
spectral weight in each orbital shows a gap at the Fermi level,
indicating a Mott insulating state in both orbitals.  The observation
of an OSPT in the intermediate region is due to the different electron
localization in the orbitals. The non-FL behavior is also present due
to weakly ordered states with AF correlations between sites and FM
correlations between orbitals in the metallic regions.

Next, using the single-site DMFT with an AF solution, we studied model
({\bf II}) and found that the critical interaction $U$ for a
metal-insulator phase transition are $U_{c_1}/t=1.9$ and
$U_{c_2}/t=2.9$ for the unfrustrated and frustrated orbitals,
respectively.  This is a clear evidence of coexistence of two phases
between the two critical $U$ values.  An OSPT is observed in this
system since the strength of electron localization in both orbitals is
different due to the unequal orbital frustration.  The AF states are a
key to the gap opening and the nature of the insulating state is that
of a band insulator.  Finally, such a minimal model may be of
relevance to understand the nature of the antiferromagnetic metallic
state in the iron-pnictide superconductors as well as the origin of
the small staggered magnetization observed in these systems since we
measure the magnetization in each orbital and find that the
magnetization only increases smoothly as a function of the interaction
strength in the antiferromagnetic metallic state.

\section{Acknowledgements}

We would like to thank H. Monien and C. Gros for useful discussions
and we gratefully acknowledge financial support
from the Deutsche Forschungsgemeinschaft through grants FOR 1346 and SPP 1458
and from the Helmholtz Association through grant HA216/EMMI.

\end{document}